\documentclass[]{pasj01}

\begin{document}
\Received{}
\Accepted{}


\title{\,Luminosity and spin-period evolution of GX~304$-$1 during
outbursts from 2009 to 2013 observed with the MAXI/GSC, RXTE/PCA,
and Fermi/GBM}

\author{
  Mutsumi \textsc{Sugizaki},\altaffilmark{1}
  Takayuki \textsc{Yamamoto},\altaffilmark{1,2}
  Tatehiro \textsc{Mihara},\altaffilmark{1}
  Motoki \textsc{Nakajima},\altaffilmark{3}
  Kazuo \textsc{Makishima},\altaffilmark{1,4}
}
\altaffiltext{1}{MAXI team, RIKEN, 2-1 Hirosawa, Wako, Saitama 351-0198}
\altaffiltext{2}{Department of Physics, Nihon University, 
  1-8-14 Surugadai, Chiyoda, Tokyo 101-8308}
\altaffiltext{3}{School of Dentistry at Matsudo, Nihon University, 
  2-870-1 Sakaecho-nishi, Matsudo, Chiba 101-8308}
\altaffiltext{4}{Department of Physics, The University of Tokyo, 
  7-3-1 Hongo, Bunkyo, Tokyo 113-0033}
\email{sugizaki@riken.jp}



\maketitle

\begin{abstract}
A report is made on the luminosity and pulse-period evolution of the
Be binary X-ray pulsar, GX 304$-$1, during a series of
outbursts from 2009 to 2013 observed by MAXI/GSC, RXTE/PCA, and
Fermi/GBM.
In total, twelve outbursts repeated by $\sim$ 132.2 days
\textcolor{black}{were observed}, which is
consistent with the X-ray periodicity of this object observed in the
1970s.
These 12 outbursts, together with those in the 1970s, were found to
all recur with a well defined period of 132.189$\pm$0.02 d, which 
can be identified with the orbital period.
The pulse period at $\sim 275$ s, obtained from the RXTE/PCA and
Fermi/GBM data, apparently exhibited a periodic modulation
synchronized with the outburst period, suggesting the pulsar orbital
motion, which is superposed on a secular spin-up trend throughout the
entire active phase.
The observed pulse-period changes were successfully represented by a
model composed of the binary orbital modulation and pulsar spin up
caused by mass accretion through an accretion disk.
The orbital elements obtained from the best-fit model, including the
projected orbital semi-major axis $a_{\rm x}\sin i \simeq 500-600$
light-s and an eccentricity $e \simeq 0.5$, are typical of Be
binary X-ray pulsars.
\end{abstract}



\section{Introduction}

X-ray binary pulsars (XBPs) are systems consisting of magnetized
neutron stars and mass-donating stellar companions. According to the
type of the companion, they are classified into several subgroups,
including Super Giant XBPs and Be XBPs as major members
(e.g. \cite{Reig2011}).
%
%
Since these neutron stars are strongly magnetized, the matter flows
from the companion are dominated by the magnetic pressure inside the
Alfven radius, and are then funneled onto the magnetic poles along the
magnetic field lines.
Since the accreting matter transfers its angular momentum at the
Alfven radius to the neutron star, the spin-up rate and the mass
accretion rate (i.e. the X-ray luminosity) of an XBP are thought to be
closely correlated (e.g. \cite{Ghosh_Lamb1979}).
The issue is relevant to the evolution scenario
of XBPs and their magnetic fields, 
and thus very important.
%
%
Although several observational results to examine the relation have
been reported so far (e.g. \cite{1996ApJ...459..288F,
  1996A&A...312..872R, 1997ApJS..113..367B}), the question is still
unsettled
because we need to select an object with a well-determined
magnetic-field strength and large luminosity swings.
Furthermore, we need to disentangle intrinsic period changes
from orbital Doppler effects.

GX 304$-$1 is a Be XBP discovered by X-ray balloon
observations in 1967 (e.g.~\cite{McClintock1971}).  The source shows
X-ray properties typical of Be/X-ray binaries; large flux
variations (\cite{Ricker1973}), the $272\pm 0.1$-s coherent pulsation
(\cite{Huckle1977}; \cite{McClintock1977}), a hard X-ray spectrum
represented by a power-law with an absorption column density $N
\rm{_{H}} \sim 1 \times 10^{22}$ cm$^{-2}$ and a photon index $\Gamma
\sim$2 up to 40 keV \citep{White1983}, and  
periodic flaring events, namely outbursts, separated by $\sim 132.5$ days
presumably attributed to the
binary orbital period \citep{PriedhorskyandTerrell1983}.
The optical counterpart was identified with a Be
star \citep{Mason1978}, which is characterized by 
double-peak H$_\alpha$ emission, strong
He$_{\rm I}$ and O$_{\rm I}$ absorptions 
\citep{Thomas1979, Parkes1980}, and photometric
variability \citep{Menzies1981}.  
Based on these line features,
\citet{Corbet1986} suggested that the circumstellar disk
is edge-on ($i \sim 90^{\circ}$).  
From the visual extension
($A_V = 6.9$ mag.), the source distance is estimated to be $2.4 \pm
0.5$ kpc \citep{Parkes1980}.
However, the binary orbital elements have not been determined, yet.

Since the last outburst was observed in the 1980s \citep{Pietsch1986}, the
source had been in a quiescent state with no significant X-ray
emission for 28 years.
Its long absence was broken by the hard X-ray
detection with INTEGRAL in 2008 June \citep{Manousakis2008}.
Since then, continuous X-ray monitoring with
the MAXI GSC 
and the Swift BAT 
has detected nine outbursts every $\sim 132.5$ d,
from 2009 November to 2012 November
\citep{Yamamoto2009, Krimm2010, Mihara2010a,
Nakajima2010, Kuhnel2010, Yamamoto2011a,
Yamamoto2011b, Nakajima2012}.
Across these multiple outbursts, the 2--20 keV MAXI GSC
intensity varied nearly by two orders of magnitude.
Suzaku, RXTE and INTEGRAL observations, performed at a bright
outburst phase, revealed a cyclotron resonance scattering feature
(CRSF) at around 54 keV \citep{Yamamoto2011pasj, Klochkov2012b}.  The
estimated surface magnetic field, 
\textcolor{black}{
$B_{\rm s}=4.7\times 10^{12} (1+z_{\rm G})$ G where $z_{\rm G}$ represents
the gravitational redshift,
}
is almost
the highest among those of XBPs of which the CRSF signature
is significantly detected.
In the 2010 August outburst,
the barycentric pulsation period was $\sim$275.37-275.46 s, 
which suggests that the pulsar was spun down by $\sim3$ s
during the 28 years of quiescence
 \citep{Devasia2011, Yamamoto2011pasj}.  
The timing analysis of RXTE data
revealed the complex pulse-profile
dependence on the energy band as well as the luminosity, and the quasi-periodic
oscillation at $\sim 0.12$ Hz \citep{Devasia2011}.
The period and pulsed flux have also been monitored by the Fermi GBM
pulsar project \citep{2013ATel.4812....1F}.
All these data thus provide us with a valuable opportunity
to study the relation between the luminosity and spin period
of the XBP.

In this paper, we analyze the X-ray outburst light curve of GX 304$-$1
during the active period from 2009 to 2013 obtained with the MAXI GSC,
and the pulse period variations derived from 
the RXTE/PCA and Fermi/GBM data.  
Our goal is to separate the orbital doppler effects
and the intrinsic pulse-period change.
The observation and data reduction are described in section
\ref{sec:observation}, while the data analysis and 
results in section \ref{sec:analysis}.  
We discuss the results in section \ref{sec:discussion}.

\section{Observation}
\label{sec:observation}

\subsection{MAXI monitoring}
\label{sec:obs_maxi}

\begin{figure}
  \begin{center}
    \includegraphics[width=8.2cm]{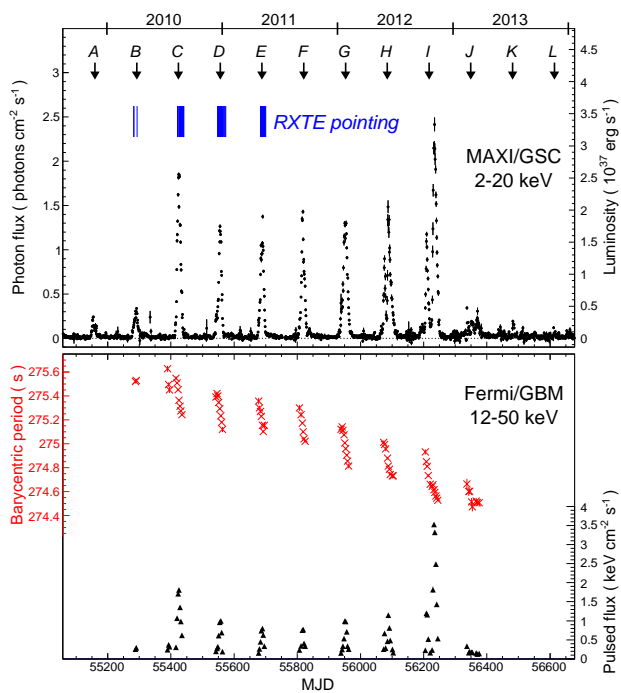}
  \end{center}
  \caption{ (Top) MAXI GSC 2--20 keV light curve of GX 304$-$1 from
    2009 August 15 (MJD 55058) to 2014 January 21 (MJD 56678) in 1-d
    time bin.  Arrows indicate the expected outburst epochs of the
    132.2-d cycle.  Twelve epochs involved in the light curve are
    named A, B, C, .., and L, as labeled.  Vertical bars in blue
    indicate the epochs of the RXTE pointing observations.  (Bottom)
    Barycentric pulse period (red cross) and pulsed 12--50 keV flux
    (black solid triangle) measured by the Fermi GBM.  }
  \label{fig:gsclc}
\end{figure}

Since the MAXI (Monitor of All-sky X-ray Image;
\cite{Matsuoka_pasj2009}) experiment onboard the ISS (International Space
Station) started its operation in 2009 August, the GSC (Gas Slit
Camera; \cite{Mihara_pasj2011}), one of the two MAXI detectors, has
been scanning almost the whole sky every 92-minute orbital cycle in
the 2--30 keV band.
We utilized archived GSC light-curve data for GX 304$-$1 
in 2--4 keV, 4--10 keV and 10--20 keV bands,
which are processed with a standard procedure \citep{sugizaki_pasj2011}
by the MAXI team 
and available from the MAXI web site\footnote{http://maxi.riken.jp}.
Figure \ref{fig:gsclc} top shows the 2--20 keV MAXI/GSC light curve 
from 2009 August to 2013 December in 1-d time bin.
It clearly reveals the recurrent outburst activity by a $\sim 132.2$ d
interval, which is consistent with the $132.5\pm 0.4$ d periodicity
found in the 1971--1972 outbursts with Vela 5B
\citep{PriedhorskyandTerrell1983}.  
Figure \ref{fig:gsclc} covers 12 epochs of the outbursts
\textcolor{black}{
predicted from the 132.2-d period cycle}, 
which we consecutively name  A, B, C, ..., and L.
The outburst I in 2012 August reached $\sim 2.4$ photons cm$^2$ s$^{-1}$
($\simeq 0.6$ Crab) in the 2--20 keV band, which is the highest among the
flaring events that have ever been observed from this source.

If the source distance of $D=2.4$ kpc, the energy spectrum of a cutoff
power-law with a photon index $\Gamma=0.35$ and an e-fold energy
$E_{\rm fold}=11$ keV \citep{Yamamoto2011pasj} are employed, 
the observed flux of 1 photon cm$^{-2}$ s$^{-1}$ in the 2--20 keV band 
corresponds to the bolometric luminosity
of $1.38\times 10^{37}$ erg s$^{-1}$ in an isotropic emission source.
We use the conversion factor to estimate the luminosity
form the MAXI/GSC data, hereafter.
The scale of the estimated luminosity is shown at the right-hand
ordinate of the light-curve (figure \ref{fig:gsclc} top).  

\subsection{Fermi GBM data}

The GBM (Gamma-ray Burst Monitor; \cite{Meegan2009}) onboard the Fermi
Gamma-Ray Space Telescope is an all-sky instrument, sensitive to
X-rays and gamma-rays with energies between 8 keV and 40 MeV.  The
Fermi GBM pulsar
project\footnote{http://gammaray.nsstc.nasa.gov/gbm/science/pulsars/}
provides results of timing analysis of a number of positively detected
X-ray pulsars, including their pulse periods and pulsed fluxes
\citep{2009arXiv0912.3847F, 2010ApJ...708.1500C}.
We utilized the archived results of GX 304$-$1 to systematically
investigate the long-term variations of the pulse timing from 2010 to
2013.

Figure \ref{fig:gsclc} (bottom) shows time variations of the
barycentric pulse period and the pulsed photon flux in the 25--50
keV band of GX 304$-$1, as made publicly available by the project.
The pulsed emission was detected over bright phases of the nine
consecutive outbursts from B to J.
%
%
The period range of 275.2--275.6 s during the outburst C is consistent
with the results of 275.37 s from the RXTE data \citep{Devasia2011} and
275.46 s from the Suzaku data \citep{Yamamoto2011pasj},
both obtained in the same outbursts.

\subsection{RXTE observations and data reduction}

The RXTE observations of GX 304$-$1 were performed on the outbursts B, C,
D, and E, 
using the Proportional Counter Array 
(PCA: \cite{Jahoda2006}) operating in 3--60 keV
and the High-Energy X-ray Timing Experiment 
(HEXTE: \cite{Rothschild1998}) providing 20--250 keV data.
As indicated in figure \ref{fig:gsclc} (top),
total 71 observations covered the outbursts C, D, and E with a
high frequency of almost once per day, each with an exposure of
0.5--18 ks.  
We analyzed the PCA data to study the pulse timing properties with 
photon statistics and time resolution better than those
available with other instruments.

Data reduction and analysis were performed using the standard RXTE
analysis tools released as a part of HEASOFT 6.14, and the CALDB files
of version 20111205 provided by NASA/GSFC.  All the data were first screened
with the standard selection criteria, that the spacecraft pointing
offset should be smaller than 0.02, the earth-limb elevation angle be larger
than 10$^{\circ}$, and the time since the last SAA passage be longer
than 30 minutes.  We used data only from the top layer of
the PCU-2 unit, which is the best calibrated among all PCUs.
In the timing analysis,
we used the reduced data with 
Good-Xenon data mode, 
Generic event-data mode (E\_125us\_64M\_0\_1s),
or Generic binned-data mode (B\_250ms\_128M\_0\_254),
which has a time resolution better than 0.25 s.

\section{Analysis}
\label{sec:analysis}

\subsection{Outburst intervals and orbital period}
\label{sec:ana_outper}

To quantify the outburst periodicity, 
we determined their peak epochs by fitting each profile with a Gaussian function.
Since some of the peaks have asymmetric profiles and hence cannot be
approximated by a single Gaussian function, the fit was performed
within a narrow span of $\sim 50$ d around each peak so that the fit is reasonably accepted.  
If outburst has multiple peaks, their peak epochs were determined 
separately.
Table \ref{tab:outpeak} summarizes the obtained epochs, fluxes,
and the luminosities estimated from the typical energy spectrum
at the peaks.

X-ray outbursts of Be X-ray binaries are known to be largely
classified into normal-type (type-I) and giant-type (type-II) ones
(e.g. review by \cite{Reig2011}).  The normal outbursts emerge near
the pulsar periastron passage and their peak luminosity reaches $\sim
10^{37}$ erg s$^{-1}$, while the giant ones may appear at any orbital
phase and can be \textcolor{black}{more} significantly luminous than the other type of events.
The regular periodicity of the observed outbursts suggests that they
are mostly categorized into the normal type.  Actually, the first
seven outbursts from A to G, with a single-peak profile, satisfy all
the normal-type conditions.
Similarly, the outbursts observed in 1971--1972 with Vela 5B also
exhibited the same periodicity and their peak fluxes were in the same
level \citep{PriedhorskyandTerrell1983}.  Therefore, they should be
categorized into the normal type, too.

\begin{figure}
  \begin{center}
    \includegraphics[width=8.2cm]{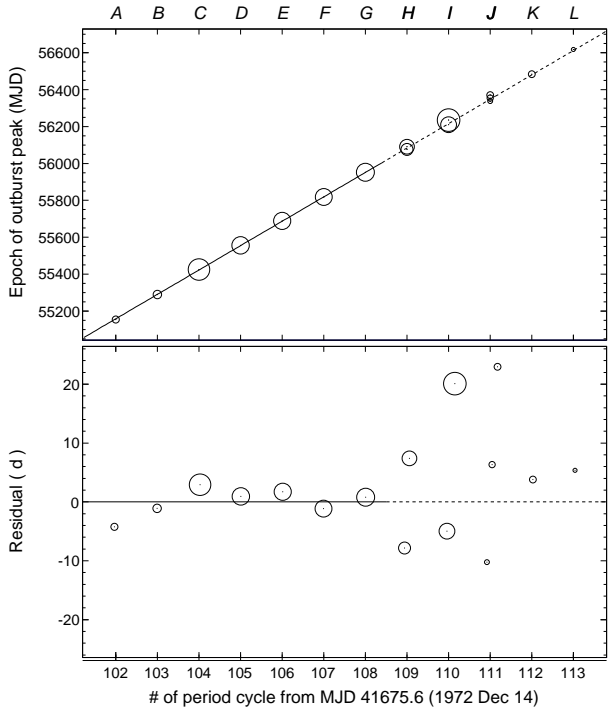}
  \end{center}
  \caption{ (Top) Epochs of outburst peaks against the number of the
    132.2-d period cycles counted from that on MJD 41675 (1972 December 14).
    The solid line represents the best-fit linear function to the data
    of cycle$=0$ at MJD 41675 and cycle$=$102--108, (Bottom) Residuals
    of the data from the best-fit linear function.  The area of the
    circles are proportional to the outburst-peak intensity in
    both panels.  }
  \label{fig:lc_phase}
\end{figure}

To refine the orbital period, we extrapolated the periodicity of the
2009--2012 outbursts (A to G) to that in 1971--1972 assuming that they
are at the same orbital phase.
The top panel of figure \ref{fig:lc_phase} shows a plot of the peak
epochs $T_{\rm peak}$ in 2009--2013 against the number $n$ of 132.2-d cycles from the
initial epoch of MJD 41675.6 (1972 December 14) in Vela 5B era.  
Thus, the epochs $T_{\rm peak}$ 
of $n=0$ (MJD 41675.6) and $n=$102--108 (outbursts A \textcolor{black}{to} G)
line up linearly, around which the data of $n=$102--108 scatter by 2.1 d
in the standard deviation.
We employed this deviation as the uncertainty in each data, and fitted
the epochs with a linear function to obtained the best period estimate
of $132.1885\pm 0.022$ and the initial peak epoch of MJD $41675.0\pm
2.2$ with 1-$\sigma$ error.
The best-fit function,
\begin{equation}
T_{\rm peak} = 41675.0 + 132.189 \times n ~~~ ({\rm MJD})
\label{equ:peakepoch}
\end{equation}
is drawn together on the data at the top panel of figure \ref{fig:lc_phase},
and the bottom panel shows the data-to-model residuals.  Their values
are listed in table \ref{tab:outpeak}.

As presented in figure \ref{fig:lc_phase} and table \ref{tab:outpeak},
we also extended the best-fit linear function to $n=113$
(outburst L).  The residuals suggest that the outburst periodicity
changed after $n=109$ (outburst H), as investigated below.

\begin{longtable}{ccrccc}
\caption{Outburst peaks of the GX 304$-$1 active period in 2009--2013 observed with MAXI GSC}\label{tab:outpeak}
\hline
OutID$^\dagger$ & Epoch & \multicolumn{2}{c}{Residual$^\ddagger$} & Peak flux$^\S$ & Luminosity$^\#$\\
          & (MJD) & \multicolumn{2}{c}{( d )} & ( photons cm$^2$ s$^{-1}$ ) & ( $10^{37}$ erg s$^{-1}$ )\\
\endfirsthead
\hline
OutID$^\dagger$ & Epoch & \multicolumn{2}{c}{Residual$^\ddagger$} & Peak flux$^\S$ & Luminosity$^\#$\\
          & (MJD) & \multicolumn{2}{c}{( d )} & ( photons cm$^2$ s$^{-1}$ ) & ( $10^{37}$ erg s$^{-1}$ )\\
\endhead
\hline
\multicolumn{6}{l}{$^*$: Second or third highest peak in a outburst cycle}\\
\multicolumn{6}{l}{$^\dagger$: Outburst ID designated by the number of period cycles (figure \ref{fig:gsclc})}\\
\multicolumn{6}{l}{$^\ddagger$: Residual of peak epoch from the best-fit period cycle of equation (\ref{equ:peakepoch})}\\
\multicolumn{6}{l}{$^\S$: Peak photon flux in 2--20 keV band in units of photons cm$^{-2}$ s$^{-1}$}\\
\multicolumn{6}{l}{$^\#$: Bolometric luminosity estimated from the 2--20 keV photon flux}\\
\endfoot
\hline
\multicolumn{6}{l}{$^*$: Second or third highest peak in a outburst cycle}\\
\multicolumn{6}{l}{$^\dagger$: Outburst ID designated by the number of period cycles (figure \ref{fig:gsclc})}\\
\multicolumn{6}{l}{$^\ddagger$: Residual of peak epoch from the best-fit period cycle of equation (\ref{equ:peakepoch})}\\
\multicolumn{6}{l}{$^\S$: Peak photon flux in 2--20 keV band in units of photons cm$^{-2}$ s$^{-1}$}\\
\multicolumn{6}{l}{$^\#$: Bolometric luminosity estimated from the 2--20 keV photon flux}\\
\endlastfoot
\hline
A     & $55154.00 \pm 0.11$ & $ -4.18$ &&  $0.198 \pm 0.005$ & 0.27\\
B     & $55289.30 \pm 0.10$ & $ -1.07$ &&  $0.275 \pm 0.004$ & 0.38\\
C     & $55425.52 \pm 0.02$ & $  2.96$ &&  $1.822 \pm 0.008$ & 2.51\\
D     & $55555.70 \pm 0.03$ & $  0.95$ &&  $1.182 \pm 0.008$ & 1.63\\
E     & $55688.66 \pm 0.03$ & $  1.73$ &&  $1.141 \pm 0.008$ & 1.57\\
F     & $55817.98 \pm 0.03$ & $ -1.14$ &&  $1.137 \pm 0.008$ & 1.57\\
G     & $55952.10 \pm 0.05$ & $  0.79$ &&  $1.266 \pm 0.009$ & 1.75\\
H$^*$ & $56075.67 \pm 0.09$ & $ -7.84$ &&  $0.564 \pm 0.025$ & 0.78\\
H     & $56090.93 \pm 0.21$ & $  7.43$ &&  $0.864 \pm 0.014$ & 1.19\\
I$^*$ & $56210.72 \pm 0.05$ & $ -4.97$ &&  $0.978 \pm 0.012$ & 1.35\\
I     & $56235.79 \pm 0.05$ & $ 20.10$ &&  $1.997 \pm 0.020$ & 2.76\\
J$^*$ & $56337.64 \pm 0.58$ & $-10.24$ &&  $0.098 \pm 0.003$ & 0.13\\
J$^*$ & $56354.21 \pm 0.23$ & $  6.33$ &&  $0.151 \pm 0.009$ & 0.21\\
J     & $56370.84 \pm 0.37$ & $ 22.96$ &&  $0.186 \pm 0.012$ & 0.26\\
K     & $56483.86 \pm 0.15$ & $  3.80$ &&  $0.177 \pm 0.009$ & 0.24\\
L     & $56617.61 \pm 1.09$ & $  5.35$ &&  $0.045 \pm 0.004$ & 0.06\\
\hline
\end{longtable}

\subsection{Evolution of outburst orbital profile}

\begin{figure*}
  \begin{center}
    \includegraphics[width=15.5cm]{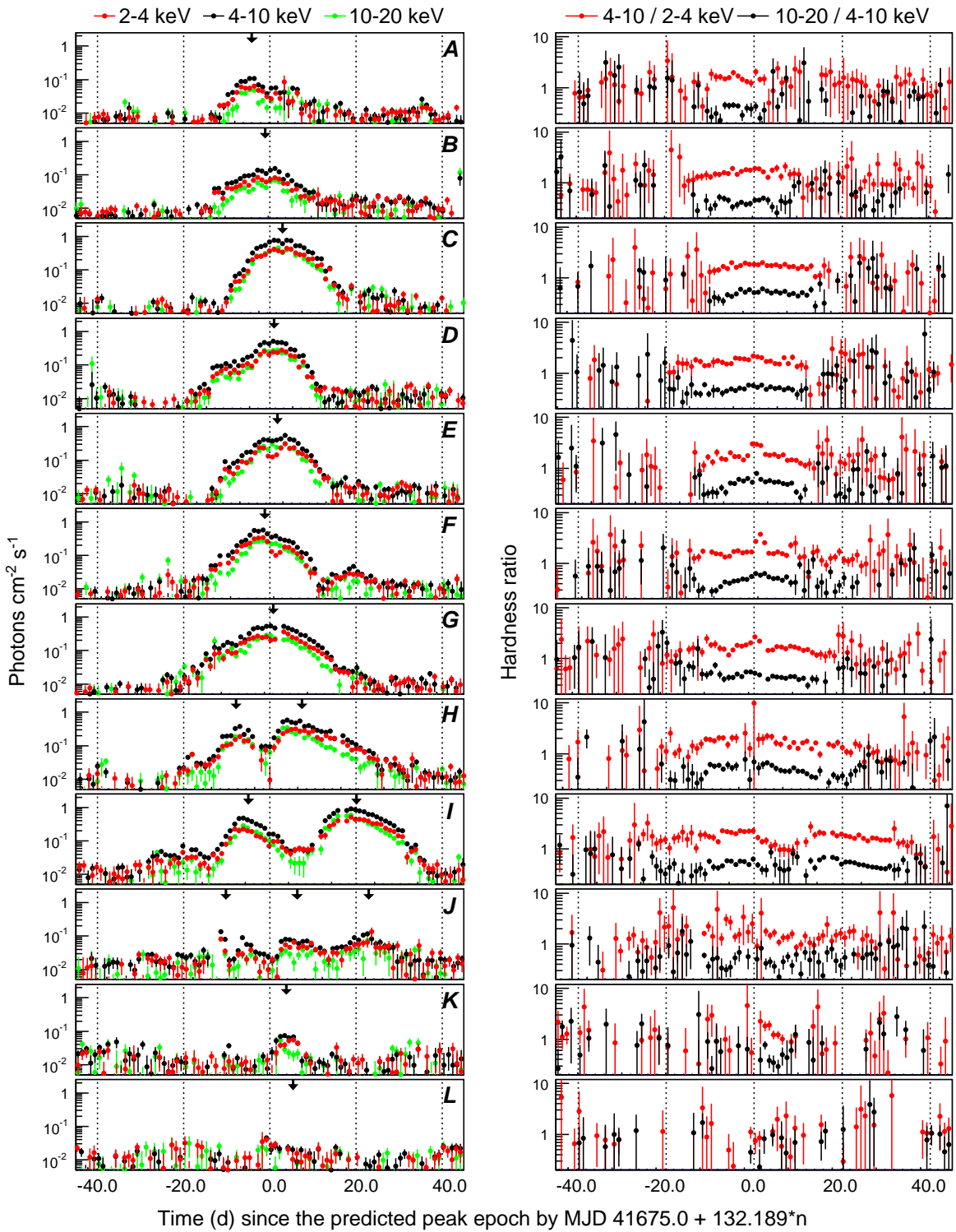}
  \end{center}
  \caption{ (Left) Light curves of the outbursts A to L, in 2--4 keV
    (red), 4--10 keV (black), and 10-20 keV (green) bands shown 
    as a function of day modulo the 132.189 d \textcolor{black}{outburst period}.  
    Arrows at the top
    of each panel indicate the peak position determined by Gaussian
    fit (table \ref{tab:outpeak}).  (Right) Time evolution of 4--10
    keV / 2--4 keV (red) and 10--20 keV / 4--10 keV (black) hardness
    ratios.}
  \label{fig:lc_orb}
\end{figure*}

Figure \ref{fig:lc_orb} left shows expanded 
three-color MAXI/GSC
light curves of the individual outbursts, as a function of 
time from the peak predicted by equation (\ref{equ:peakepoch}).
The hardness ratios are presented in the right panels.

The light curves of the outbursts A to G in figure \ref{fig:lc_orb} (left)
reconfirm that their peak epoch agrees, within a scatter of 2.1 d
with the prediction of the best-fit cycle.
However, these profiles vary considerably from cycle to cycle.
While the outburst D has a slight enhancement over the
symmetrical profile \textcolor{black}{prior} to the peak, the outbursts F and G have a
subsequent small peak after the main peak.

As expected from figure \ref{fig:gsclc}, the outbursts H and I, the
latter being the brightest, clearly deviate in figure
\ref{fig:lc_phase} and figure \ref{fig:lc_orb} form equation
(\ref{equ:peakepoch}), because they both show a double-peaked profiles.
This suggests that these two, or at least outburst I, are possibly the
giant outbursts.
In the outburst J, no distinct main peak is seen and a low-level
activity continued for 40 days.

During bright outburst phases, the two hardness ratios in figure
\ref{fig:lc_orb} (right) slightly change with a positive correlation
to the flux.  In addition, the (4--10 keV / 2--4 keV) hardness ratio
exhibits sharp increases at the peak in the outbursts E, F, and G.
They correspond to dips in the softest light curve (red), and thus can
be attributed to an increase of the absorption column density.

\subsection{Pulse period changes}
\label{seq:pulsesearch}

\begin{figure*}
  \begin{center}
    \includegraphics[width=14.cm]{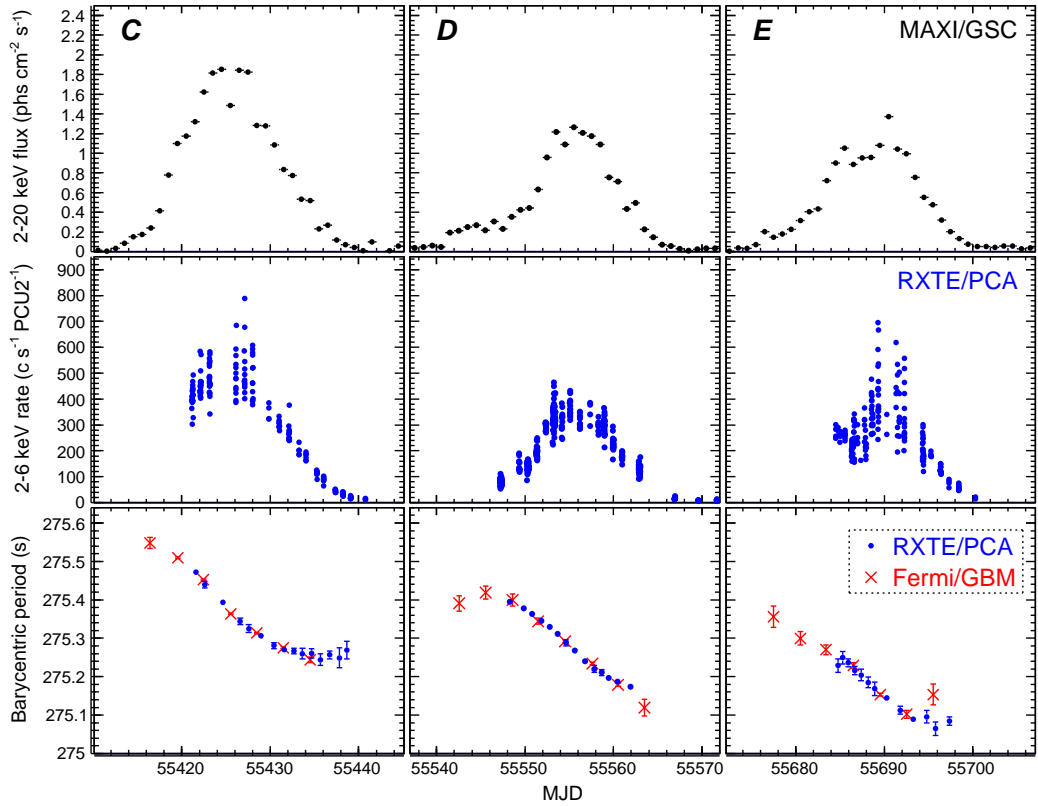}
  \end{center}
  \caption{ Details of the outbursts C (left), D (center), and E (right),
    (Top) MAXI/GSC light curves in the 2--20 keV band, in 1-d time bin.
    (Middle) THe 2-6 keV RXTE/PCA count rate per 272-s time bin 
    during the pointing observations.
    (Bottom) Barycentric pulse periods obtained from the RXTE/PCA data
    (blue dot) and the Fermi/GBM pulsar archive data (red cross)
    All vertical error bars represent the 1-$\sigma$ uncertainty.
  }
  \label{fig:pulsep}
\end{figure*}

To investigate the change in the pulse period with a better precision 
and sensitivity than those available with the Fermi GBM
shown in red in figure \ref{fig:gsclc} (bottom),
we analyzed the RXTE PCA data, which samples 
the outbursts C, D, and E on the almost daily basis.  
The data for the outburst C has been already analyzed by \citet{Devasia2011}
and the results revealed a complex pulse profile
that depends on the energy and the luminosity.
We performed the timing analysis taking account of these properties as follows.
Figure \ref{fig:pulsep} top and middle panels show
the MAXI/GSC 2--20 keV light curves in 1-d time bin, and
those of the 2--6 keV RXTE/PCA count rates per 272 s,  
approximately equal to the pulse period.
The large variability in the PCA count rate around the outburst peaks
represent the variation component other than the 275-s pulsation,
as \citet{Devasia2011} reported in the outburst C.

We performed the pulse period search by the folding method 
using  {\tt efsearch} in XRONOS,
for the PCA data after the barycentric correction was 
applied with {\tt faxbary} in FTOOL.
To optimize the period determination accuracy,
the folding analysis was carried out by combining data taken
within each interval up to 2 days. 
The energy band was limited to 2--6 keV because the sharp dip structure 
in the pulse profile below 8 keV \citep{Devasia2011}.
%
The errors on the obtained pulse period 
was estimated by the method of \citet{1996A&AS..117..197L} 
using the folded pulse profile with 64 phase bins.

Figure \ref{fig:pulsep} (bottom) show the obtained pulse period against the
observation epochs, where the results of the Fermi GBM pulsar project,
the same as in figure \ref{fig:gsclc}, are plotted together. 
The RXTE/PCA and the Fermi/GBM results are consistent with each other.

\subsection{Determining the binary orbital elements}
\label{sec:orbitalsolution}

\begin{figure*}
  \begin{center}
    \includegraphics[width=11.cm]{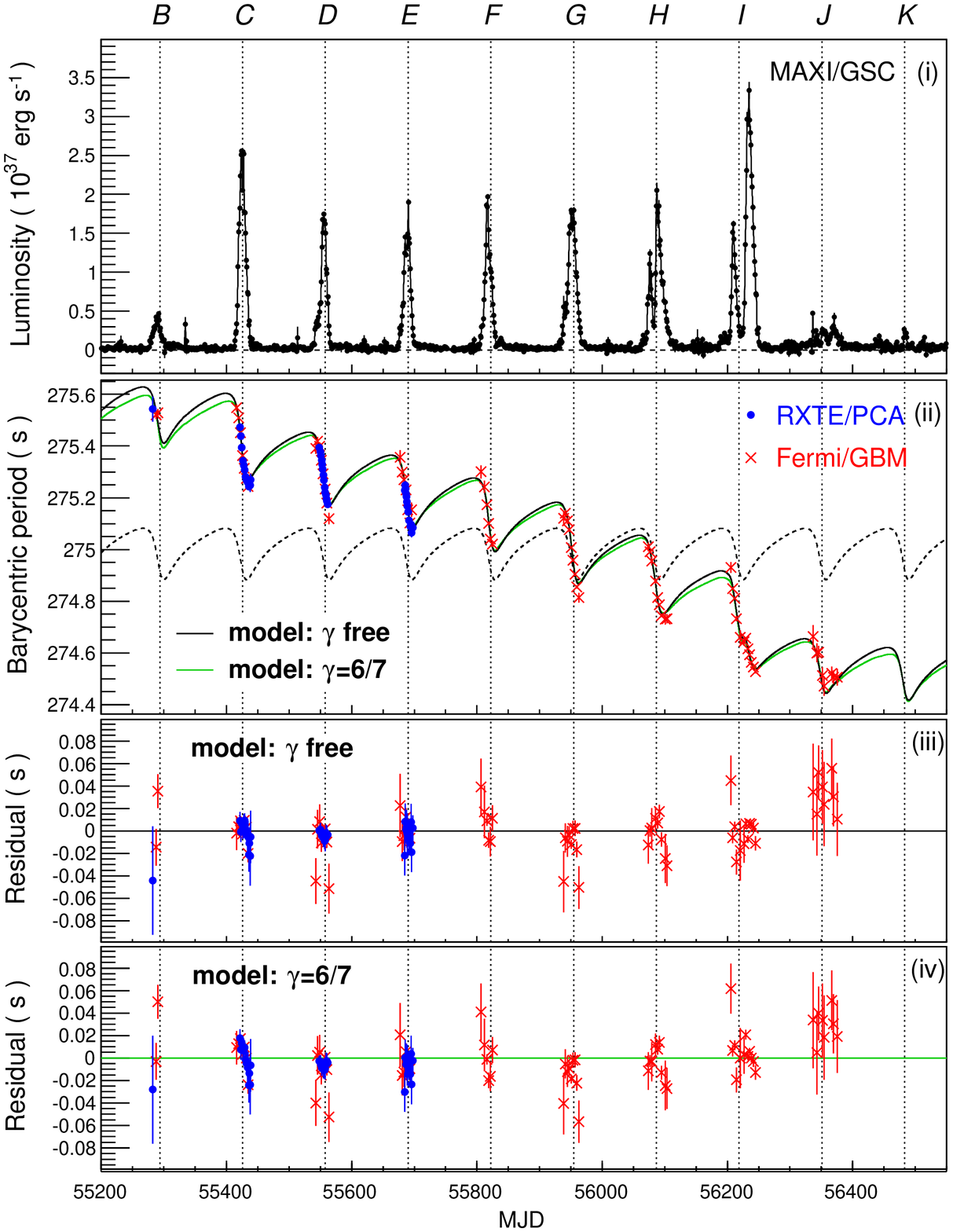}
  \end{center}
  \caption{ 
    (i) Bolometric luminosity estimated from the 2--20
    MAXI/GSC light curve in 1-d time bin.  
    (ii) Barycentric pulse
    period measured with the RXTE/PCA (blue) and the Fermi/GBM (red),
    fitted with equations (\ref{equ:deltapmodel})-(\ref{equ:pobs}).
    Black and green solid lines represent 
    the best-fit model with free $\gamma$ and that with $\gamma=6/7$, respectively.
    A dashed line represents the period modulation due to the orbital Doppler effect.
    (iii) Residuals of the data from the best-fit model with free $\gamma$.
    (iv) Residuals from the $\gamma=6/7$ model.
  }
  \label{fig:pulsehist}
\end{figure*}

The pulse-period evolution in figures \ref{fig:gsclc} and
\ref{fig:pulsep} is characterized by two distinct effects. 
One is
cyclic modulation synchronized with the 132.2 d outburst period,
considered as due to the orbital Doppler effects.  The other is a
secular spin-up trend over the 3 years, 
obviously due to the accretion torque.
Thus, the observed pulse-period evolution is 
considered to be a complex composite of
these two effects. 
Since both of them are supposed to depend on the orbital phase, 
it is not straightforward to separate them.
Hence, 
we construct a semi-empirical model representing the two effects,
and then fit it to the data, in an attempt to 
simultaneously determine the
orbital elements and the 
luminosity-dependent
intrinsic pulse period change.


Be XBPs are known to exhibit spin-up episodes during
bright outburst phases (e.g. \cite{1997ApJS..113..367B}). This is
naturally explained by an increase in the accretion rate, and 
the associated transfer of the angular momentum to the neutron star via
disc-magnetosphare interactions.
In the classical theoretical model by \citet{Ghosh_Lamb1979}, 
the pulsar spin-up rate $-\dot{P}_{\rm GL}$ (s yr$^{-1}$) is given by
\begin{equation}
-\dot{P}_{\rm GL}  =  
5.0 \times 10^{-5} \mu_{30}^{\frac{2}{7}} n(\omega_{\rm s})  
R_{\rm NS6}^{\frac{6}{7}} 
M_{\rm NS\odot}^{-\frac{3}{7}}
I_{45}^{-1} P_{\rm spin}^2 L_{37}^{\frac{6}{7}} \label{equ:pdot}
\end{equation}
where 
$\mu_{30}$, $R_{\rm NS6}$, $M_{\rm NS\odot}$, $I_{45}$, $P_{\rm spin}$, 
and $L_{\rm 37}$ 
are the magnetic dipole moment of the neutron star
in units of $10^{30}$ G cm$^{3}$, 
radius in $10^6$ cm, 
mass in $M_\odot$, 
moment of inertia in $10^{45}$ g cm$^2$,
spin \textcolor{black}{period} in s, 
and the luminosity in $10^{37}$ erg s$^{-1}$, respectively,
while $n(\omega_{\rm s})$ is a dimensionless torque that
depends on the fastness parameter $\omega_{\rm s}$,
defined by the ratio of the angular velocity of the pulsar spin
to that of the Kepler rotation of accreting matter
at around the Alfven radius.

In the application of \textcolor{black}{equation} (\ref{equ:pdot}),
GX 304$-$1 is particularly suited for several reasons.
%
\textcolor{black}{
First, the value of $\mu_{30}$ can be accurately estimated from the
CRSF detection at 54 keV, which implies the surface magnetic field,
$B_{\rm s}=4.7\times 10^{12} (1+z_{\rm G})$ G
\citep{Yamamoto2011pasj}.
If the canonical neutron-star mass $M_{\rm NS}=1.4 M_{\odot}$ 
and radius $R_{\rm NS}=10^6$~cm are assumed, the gravitational redshift $z_{\rm G}$
is 0.3, and then $\mu_{30}$ is calculated as
\begin{equation}
\mu_{30} \approx \frac{1}{2} B_{\rm s} R_{\rm NS}^3 =  3.1 ~~~ (10^{30}~{\rm G~ cm^3})
\end{equation}
\citep{1983ApJ...265.1036W}.
}
Second, a continuous record of $L_{\rm 37}$ 
is available from the MAXI GSC light curve; 
we calculate it with
the conversion factor as described in
section \ref{sec:obs_maxi}.
Furthermore, $n(\omega_{\rm s})\simeq 1.4$ can be regarded
approximately as a constant from the long spin period, $P_{\rm spin}\simeq 275$ s, 
such that $\omega_s\ll 1$ \citep{Ghosh_Lamb1979}.
\textcolor{black}{
We also employ $I_{45}$ calculated from its approximate
equation with $M_{\rm NS}$ and $R_{\rm NS}$
given by \citet{1994ApJ...424..846R} as 
}
\begin{equation}
\textcolor{black}{
I_{45}\approx 0.21 \frac{M_{\rm NS} R_{\rm NS}^2}{1-2GM_{\rm NS}/R_{\rm NS}c^2} = 1.0 ~~~ (10^{45}~{\rm g~ cm^2}).
}
\end{equation}
Combining these pieces of information,
equation (\ref{equ:pdot})
yields
\begin{equation}
\textcolor{black}{
-\dot{P}_{\rm GL}  \simeq 1.7\times 10^{-2} L_{37}^{\frac{6}{7}} ~~~ {(\rm s~ d^{-1})}.
}
\label{equ:pdot2}
\end{equation}

Equation (\ref{equ:pdot}) implies that the spin-up rate $-\dot{P}_{\rm spin}$
follows the luminosity $L$ 
as $-\dot{P}_{\rm spin}\propto L^{\frac{6}{7}}$.  
The relation has been calibrated with actual X-ray data
\citep{1996A&A...312..872R, 1997ApJS..113..367B}.
Besides this, the comparison of absolute spin-up rates with
equation (\ref{equ:pdot}) has been hampered by a large
uncertainty in the bolometric luminosity correction,
which is in turn due to beaming effects 
(e.g. \cite{1997ApJS..113..367B,  2002ApJ...570..287W}).
We hence employ a spin-up model expressed by
$-\dot{P}_{\rm spin} = \alpha L_{37}^\gamma$
in which the power-law index and
the constant coefficient $\alpha$
are treated as free parameters.


XBPs are also known to spin down during the quiescence due to the
propeller effects.  In fact, GX 304$-$1 exhibited a spin down by $\sim
3$ s for the 28 years of quiescence from 1980 to 2008, with an average
period derivative of $\dot{P}\simeq 3\times 10^{-9}$ s s$^{-1}$.  The
rate, though much smaller than the spin-up rate during bright phases (
$\dot{P}\sim -10^{-7}$ s s$^{-1}$), may not be negligible.
We accounted its contribution 
 with a constant spin-down parameter, $\beta$, 
added to $\dot{P}_{\rm spin}$ as an offset.

By using the spin-up and spin-down models discussed above, 
the intrinsic 
\textcolor{black}{pulsar-spin} period $P_{\rm spin}(t)$ is expressed as
\begin{eqnarray}
P_{\rm spin}(t) &=& P_0 + \int_{\tau_0}^t \dot{P}_{\rm spin}(\tau) d\tau \\
&=& P_0 + \int_{\tau_0}^t  \left\{ -\alpha  L_{37}^{\gamma}(\tau) +\beta\right\}d\tau 
\label{equ:deltapmodel}
\end{eqnarray}
where we set the time origin $\tau_0$ at the periastron passage in 
the outburst C and denote the pulse period at $\tau_0$ as $P_0=P_{\rm spin}(\tau_0)$.

The period modulation due to the binary motion, can be calculated 
using the binary orbital elements, which consists of the orbital period $P_{\rm B}$,
the eccentricity $e$, the projected semi-major axis $a_{\rm x}\sin i$, 
the argument of the periastron $\omega_0$,
and the epoch of periastron passage which we identified with $\tau_0$ defined above.
The pulsar orbital velocity $v_{\rm l}(t)$ along the line of sight is given as
\begin{equation}
v_{\rm l}(t)
= \frac{2\pi a_{\rm x} \sin i}{ P_{\rm B} \sqrt{1-e^2}} \left\{\cos\left(\nu(t)+\omega_0\right)+e\cos\omega_0\right\}
\label{equ:vl}
\end{equation}
where $\nu(t)$ is a parameter called ``true anomaly'' describing the
motion on the elliptical orbit, and calculated from the Kepler's
equation.
The observed barycentric period, $P_{\rm obs}(t)$, is then expressed by
\begin{eqnarray}
P_{\rm obs}(t) &=& P_{\rm spin}(t) \left(1+\frac{v_{\rm l}(t)}{c}\right)\left(1-\frac{v^2(t)}{c^2}\right)^{-\frac{1}{2}} \nonumber \\
&\simeq& P_{\rm spin}(t) \left(1+\frac{v_{\rm l}(t)}{c}\right).
\label{equ:pobs}
\end{eqnarray}

We applied the model $P_{\rm obs}(t)$ consisting of equations
(\ref{equ:deltapmodel})-(\ref{equ:pobs}) 
to the observed barycentric periods obtained 
from the RXTE/PCA timing analysis in section \ref{seq:pulsesearch}
and Fermi/GBM archival products.
The model 
involves nine free parameters; $P_0$, $\alpha$, $\beta$, and $\gamma$
in equation (\ref{equ:deltapmodel}), and orbital elements in equation
(\ref{equ:vl}).  
\textcolor{black}{ 
We utilized the luminosity calculated from the MAXI/GSC 2--20
keV light curve in 1-d time bin, as described in section
\ref{sec:obs_maxi}.
}
\textcolor{black}{ 
%
%
%
The errors associated with the $L_{37}$ determinations 
propagate into those of equation (\ref{equ:deltapmodel}), 
but this gives 1-$\sigma$ uncertainties only by 
$\sim 0.003$ s, which is smaller than those in the individual
pulse-period measurements, $\sim 0.01$ s.
}
We fixed the orbital period $P_{\rm B}$ at 132.189 d
derived from the outburst cycle over 28 years in section
\ref{sec:ana_outper}, and examine the fit with $\gamma$ fixted at the
theoretical value of 6/7 or free.

\begin{figure}
  \begin{center}
    \includegraphics[width=7.5cm]{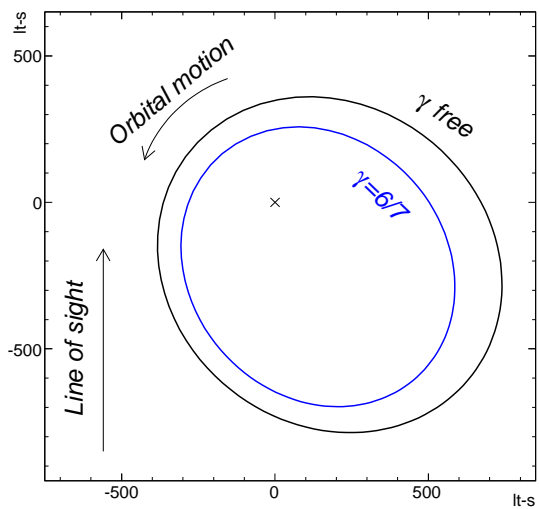}
  \end{center}
  \caption{
    Pulsar binary orbits obtained in the
    two period-change models with $\gamma$ free (black) and $\gamma=6/7$ (blue),
    projected on a plane including the line of sight.
  }
  \label{fig:pulsarorbit}
\end{figure}

The model has been found to represent the data well.  
Figure \ref{fig:pulsehist} shows the results of those model fits
(with $\gamma=6/7$ and $\gamma$ free), and 
table \ref{tab:binaryparam} summarizes the obtained best-fit
parameters.
\textcolor{black}{
The obtained epoch of the periastron of the outburst C,
$\tau_0\simeq 55425$~(MJD),
agrees with that at the outburst flux peak (table \ref{tab:outpeak})
within the uncertainty of the peak-epoch periodicity, $\sim 2.1$ d.
}
Comparing the two fit results, the fit with $\gamma$ free 
\textcolor{black}{has a $\chi^2$ lower by $108.3$}
for a decrease of only by one in the degree of freedom (dof).
In figure \ref{fig:pulsarorbit}, pulsar binary orbits 
suggested in the two fitting models,
are illustrated.

\begin{table}
\tbl{Best-fit orbital parameters}{
  \begin{tabular}{lcc} 
    \hline
    & \multicolumn{2}{c}{Model}\\
    Parameter       &  $\gamma=6/7$ & $\gamma$: free \\   
    \hline
    $P_{\rm B}$ ( d )                   & \multicolumn{2}{c}{132.189 (fix)} \\
    $a_{\rm x}\sin i$ ( lt-s )          &  $498 \pm 6$ & $601 \pm 38$ \\                           
    $e$                                &   $0.524 \pm 0.007$ & $0.462 \pm 0.019$ \\                
    $\tau_0$ ( MJD )                   &   $55425.0200 \pm 0.0010$ & $55425.6 \pm 0.5$ \\    
    $P_0$ ( s )                        & $275.4441 \pm 0.0008$ & $275.459 \pm 0.008$ \\        
    $\omega_0$ ( degree )              & $122.5 \pm 0.4$ & $130.0 \pm 4.4$ \\                    
\textcolor{black}{$\alpha/0.017$}        &  \textcolor{black}{$0.2989 \pm 0.0004$} & \textcolor{black}{$0.2467 \pm 0.0010$}\\            
    $\beta$  ( $10^{-9}$s\, s$^{-1}$ )  & $3.27 \pm 0.02$ & $< 0.23$ \\                    
    $\gamma$                           &  $6/7$ (fix) & $1.243 \pm 0.031$ \\                
    $f_{\rm M}$ ( $M_\odot$ )            & $7.6 \pm 0.3$ & $13.3 \pm 2.5$ \\                       
    $\chi^2$ / dof                     &   276.3 / 103 & 168.0 / 102\\
    \hline
  \end{tabular}
}\label{tab:binaryparam}
\begin{tabnote}
All errors represent 1-$\sigma$ confidence limit of the statistical uncertainty.
\end{tabnote}
\end{table}

\section{Discussion}
\label{sec:discussion}

\subsection{Lumonosity - spin-up relation}

As a result of the model fit to the pulse-period evolution, we
successfully calibrate the relation between
the the luminosity $L$ and the spin-up rate, 
\textcolor{black}{$-\dot{P}_{\rm spin}$},
which is considered to reflect the accretion manner.
According to the model of \citet{Ghosh_Lamb1979},
the power-law index $\gamma$ becomes 6/7 in a disk accretion and 1 in a wind accretion.
The best-fit $\gamma$ obtained from the model of $\gamma$ free was $1.24\pm 0.03$, 
which is significantly larger than $6/7\simeq 0.86$ in
the disk accretion.
%
%
%
Meanwhile, the present result agrees with those of other typical Be XBPs, 
A 0535$+$26 \citep{1996ApJ...459..288F, 2012ApJ...754...20C},
EXO 2030$+$375 \citep{1996A&A...312..872R, 2002ApJ...570..287W},
and 2S 1417-624 \citep{1996A&AS..120C.209F, 2004MNRAS.349..173I}.
As \citet{2002ApJ...570..287W} suggested, 
the result of $\gamma \gtsim 1$ may imply
the effect of either the spin down 
or the wind accretion during the low luminosity periods.
In fact, the fit with $\gamma=6/7$ yielded the significant
spin-down of $\beta = 3.27 \pm 0.02 \times 10^{-9}$ s s$^{-1}$ during
the outburst intermissions.  The rate is consistent with the average
period derivative of $\sim 3 \times 10^{-9}$ s s$^{-1}$ during the
1980-2008 quiescence.

Comparing the two model-fit results ($\gamma=6/7$ and $\gamma$ free),
goodness of fits ($\chi^2$ / dof) certainly prefers the one with
$\gamma$ free.  However, neither is acceptable statistically, As seen
in figure \ref{fig:pulsehist}, the difference between the two models
is mostly in the outburst intermission in which the period data are
not available.  Therefore, it is hard to conclude which model better
represents the real situation, from the present result.
\textcolor{black}{
The problem will be solved if the pulse period 
throughout the entire orbital phase is obtained.
}

From the best-fit parameters, 
we also obtained the normalization factor of the absolute spin-up
rate, $\alpha$, compared to that predicted in the model of
\citet{Ghosh_Lamb1979}, which is $0.017$ s d$^{-1}$ as derived in
equation (\ref{equ:pdot2}) in this source.
The ratio, $\alpha/0.017\simeq 0.3$ in both the two model ($\gamma$
free and $\gamma=6/7$), is significantly smaller than unity, like in
some other Be XBPs \citep{2002ApJ...570..287W}.
\textcolor{black}{
This discrepancy is reasonably explained by the uncertainty 
in the luminosity estimate.
We calculate the luminosity from the observed fluxes sampled
by MAXI/GSC scan survey in 1-d time bin assuming that the source 
emission is isotropic, as described in section \ref{sec:obs_maxi}.
Because the observed flux is
pulsating, the source emission is not completely isotropic.  
The obtained result corresponds to the beaming effect by a factor $\sim 3$.
}

\subsection{Binary properties as Be/X-ray binary pulsars}

From the obtained orbital elements, the mass function $f_{\rm M}$ is 
calculated as
\begin{equation} 
f_{\rm M} 
= \frac{(M_{*} \sin i)^3}{(M_{*}+M_{\rm NS})^2}
= \frac{4\pi^2}{G}\frac{(a_{\rm x} \sin i)^3}{P_{\rm B}^2},
\end{equation}
where $G$ is the gravitational constant, and $M_{*}$ is the mass of the
companion star.  
Table \ref{tab:binaryparam} yields
$f_{\rm M}=7.6\pm 0.3M_\odot$ if $\gamma=6/7$, 
and $13.3\pm 2.5 M_\odot$ if $\gamma$ is left free. 
Assuming the mass
of the pulsar to be $M_{\rm NS} = 1.4 M_\odot$, the condition of 
$\sin i \leq 1.0$ constrains the companion mass as $M_{*}\gtrsim 9.9
M_\odot$ for $f_{\rm M}=7.6 M_\odot$ and $\gtrsim 15.7 M_\odot$ for
$13.3 M_\odot$.  This agrees with the typical Be-star mass.
The system is actually suggested to be edge-on ($i\sim 90^\circ$)
from the emission and absorption lines in the optical spectra \citep{Corbet1986}.

The obtained orbital eccentricity of $e\simeq 0.5$ is typical of a
Be/XBP subgroup that is characterized by a long spin period $P_{\rm
  spin}\gtsim 100$ s and a long orbital period $P_{\rm B}\gtsim 100$ d
\citep{2011Natur.479..372K}.
The value of $e\simeq 0.5$ and $B_{\rm s}=4.7\times 10^{12} (1+z_{\rm G})$ G
of GX 304$-$1 are in line with the positive correlation
found by \citet{2014PASJ..tmp...44Y}
between these two quantities.

\subsection{Evolution of outburst light-curve profiles and Be disk}

The outburst orbital profile is considered to reflect the spatial
extent of the circumstellar disk around the Be star on the pulsar
orbit.  During the 3.5 years of active period from 2009 October to 2012
November, the profile changed through a series of normal outbursts
(A to G), giant outbursts with a large peak luminosity (H and I), 
and decay phase (J to L).
This evolution behavior vary much resembles those of other Be X-ray
binaries, including A 0535$+$26 \citep{2012ApJ...754...20C,
  2014PASJ...66....9N}, and GRO J1008-57 \citep{2013A&A...555A..95K,
  2014PASJ..tmp...44Y}.  Therefore, the same physical mechanism
is considered to be working in those systems.
To explain this, a couple of scenarios, including the global one-armed 
oscillation and the precession of the Be disk, have been proposed
(\cite{2014PASJ...66....9N} and references therein).
The detailed modeling is out of the scope of the present paper.

\bigskip

We thank all the ISS/MAXI team members for their dedicated work
to enable the science mission operation.
We also thank the Fermi/GBM pulsar project 
for providing the useful results to the public.
This research has made use of RXTE data
obtained from the High Energy Astrophysics Science Archive Research
Center (HEASARC) provided by NASA/GSFC; 
The present work is partially supported by the Ministry of Education,
Culture, Sports, Science and Technology (MEXT), Grant-in-Aid No.
24340041 and 25400239.



\begin{thebibliography}{}




\bibitem[Bildsten et al.(1997)]{1997ApJS..113..367B} 
Bildsten, L., et al.\ 1997, \apjs, 113, 367 


\bibitem[Camero-Arranz et al.(2010)]{2010ApJ...708.1500C} 
Camero-Arranz, A., Finger, M.~H., Ikhsanov, N.~R., Wilson-Hodge, C.~A., 
\& Beklen, E.\ 2010, \apj, 708, 1500 

\bibitem[Camero-Arranz et al.(2012)]{2012ApJ...754...20C} 
Camero-Arranz, A., et al.\ 2012, \apj, 754, 20 


\bibitem[Corbet et al.(1986)]{Corbet1986} 
Corbet, R.~H.~D., et al.\ 1986, \mnras, 221, 961 

\bibitem[Devasia et al.(2011)]{Devasia2011}
Devasia, J., James, M., Paul, B., \& Indulekha, K.\ 2011, \mnras, 417, 348 


\bibitem[Finger et al.(2009)]{2009arXiv0912.3847F} 
Finger, M.~H., et al.\ 2009, arXiv:0912.3847


\bibitem[Finger et al.(1996)]{1996A&AS..120C.209F} 
Finger, M.~H., Wilson, R.~B., \& Chakrabarty, D.\ 1996, \aaps, 120, 209

\bibitem[Finger et al.(1996)]{1996ApJ...459..288F} 
Finger, M.~H., Wilson, R.~B., \& Harmon, B.~A.\ 1996, \apj, 459, 288 



\bibitem[Finger \& Jenke(2013)]{2013ATel.4812....1F} Finger, M.~H., \&
  Jenke, P.~A.\ 2013, The Astron. Telegram, 4812, 1




\bibitem[Ghosh \& Lamb(1979)]{Ghosh_Lamb1979}
  Ghosh, P., \& Lamb, F.~K.\ 1979, \apj, 234, 296 

\bibitem[{\.I}nam et al.(2004)]{2004MNRAS.349..173I} {\.I}nam, S.~{\c C}., 
Baykal, A., Matthew Scott, D., Finger, M., 
\& Swank, J.\ 2004, \mnras, 349, 173 


\bibitem[Huckle et al.(1977)]{Huckle1977} 
Huckle, H. E., Mason, K. O.,  White, N. E., Sanford, P. W., Maraschi, L., Tarenghi, M., Tapia, S.,
  1977, \mnras, 180, 21P

\bibitem[Jahoda et al.(2006)]{Jahoda2006} 
  Jahoda, K., Markwardt,
  C. B., Radeva, Y., Rots, A. H., Stark, M. J., Swank, J. H.,
  Strohmayer, T. E., \& Zhang, W.\ 2006, \apjs, 163, 401


\bibitem[Klochkov et al.(2012b)]{Klochkov2012b} 
Klochkov, D., et al.\ 2012, \aap, 542, L28

\bibitem[Knigge et al.(2011)]{2011Natur.479..372K} Knigge, C., Coe, M.~J., 
\& Podsiadlowski, P.\ 2011, \nat, 479, 372 


\bibitem[Krimm et al.(2010)]{Krimm2010} 
Krimm, H.~A., et al.\ 2010, The Astron. Telegram, 2538, 1 

\bibitem[K{\"u}hnel et al.(2010)]{Kuhnel2010} 
K{\"u}hnel, M., et al.\ 2010, The Astron. Telegram, 3087, 1

\bibitem[K{\"u}hnel et al.(2013)]{2013A&A...555A..95K} K{\"u}hnel, M.,
  M{\"u}ller, S., Kreykenbohm, I., et al.\ 2013, \aap, 555, A95


\bibitem[Larsson(1996)]{1996A&AS..117..197L} Larsson, S.\ 1996, \aaps,
  117, 197



\bibitem[Manousakis et al.(2008)]{Manousakis2008}
  Manousakis \etal, 2008, Astron. Telegram, 1613

\bibitem[Mason et al.(1978)]{Mason1978} Mason, K. O., Murdin, P. G.,
  Parkes, G. E. Visvanathan, N., 1978, \mnras, 184, 45P

\bibitem[Matsuoka et al.(2009)]{Matsuoka_pasj2009} 
  Matsuoka, M., et al. \ 2009, \pasj, 61, 999

\bibitem[Meegan et al.(2009)]{Meegan2009} 
  Meegan, C., et al.\ 2009, \apj, 702, 791 

\bibitem[McClintock et al.(1971)]{McClintock1971} 
  McClintock, J. E., Ricker, G. R., Lewin, W. H. G., 1971, \apjl, 166, L73

\bibitem[McClintock et al.(1977)]{McClintock1977}
  McClintock, J. E., Rappaport, S. A. Nugent, J. J., Li, F. K., 1977, \apjl, 216, L15

\bibitem[Menzies et al.(1981)]{Menzies1981}
  Menzies, J., 1981, \mnras, 195, 67P





\bibitem[Mihara et al.(2004)]{2004ApJ...610..390M} Mihara, T., Makishima, 
K., \& Nagase, F.\ 2004, \apj, 610, 390 

\bibitem[Mihara et al.(2010a)]{Mihara2010a} 
  Mihara, T., et al.\ 2010a, The Astron. Telegram., 2779, 1

\bibitem[Mihara et al.(2010b)]{Mihara2010b} 
  Mihara, T., et al.\ 2010b, The Astron. Telegram., 2796, 1

\bibitem[Mihara et al.(2011)]{Mihara_pasj2011} 
  Mihara, T., \etal \  2011, \pasj, 63, 623


\bibitem[Nakajima et al.(2010)]{Nakajima2010} 
  Nakajima, M., et al.\ 2010, The Astron. Telegram, 3075, 1

\bibitem[Nakajima et al.(2012)]{Nakajima2012} 
  Nakajima, M., et al.\ 2012, The Astron. Telegram, 4420, 1


\bibitem[Nakajima et al.(2014)]{2014PASJ...66....9N} 
Nakajima, M., et al.\ 2014, \pasj, 66, 9 


\bibitem[Parkes et al.(1980)]{Parkes1980}
  Parkes, G.E., Murdin, P.G., Mason, K.O., 1980, \mnras, 190, 537

\bibitem[Pietsch et al.(1986)]{Pietsch1986}
  Pietsch, E., Collmar, W., Gottwald, M., Kahabka, P., \"{O}gelman, H., 1986, \aap, 163, 93



\bibitem[Priedhorsky \& Terrell (1983)]{PriedhorskyandTerrell1983}
  Priedhorsky, W. C., Terrell, J., 1983, \apj, 273, 709

\bibitem[Reig (2011)]{Reig2011} 
  Reig, P.\ 2011, \apss, 332, 1

\bibitem[Ravenhall \& Pethick(1994)]{1994ApJ...424..846R} Ravenhall,
  D.~G., \& Pethick, C.~J.\ 1994, \apj, 424, 846


\bibitem[Reynolds et al.(1996)]{1996A&A...312..872R} 
Reynolds, A.~P., et al.\ 1996, \aap, 312, 872 


\bibitem[Ricker et al.(1973)]{Ricker1973} 
  Ricker, G.~R., McClintock,
  J.~E., Gerassimenko, M., \& Lewin, W.~H.~G.\ 1973, \apj, 184, 237

\bibitem[Rothschild et al.(1998)]{Rothschild1998} 
  Rothschild, R.~E., et al.\ 1998, \apj, 496, 538 



\bibitem[Sugizaki et al.(2011)]{sugizaki_pasj2011} 
  Sugizaki, M. et al.\ 2011, \pasj, 63, 635



\bibitem[Thomas et al.(1979)]{Thomas1979}
  Thomas, R.M., Morton, D.C., Murdin, P.G., 1979, \mnras, 188, 19



\bibitem[Wasserman \& Shapiro(1983)]{1983ApJ...265.1036W} Wasserman,
  I., \& Shapiro, S.~L.\ 1983, \apj, 265, 1036


\bibitem[White et al.(1983)]{White1983} 
  White, N., Swank, J., \& Holt,
  S. S., 1983, \apj, 270, 711

\bibitem[Wilson et al.(2002)]{2002ApJ...570..287W} 
  Wilson, C.~A.,
  Finger, M.~H., Coe, M.~J., Laycock, S., \& Fabregat, J.\ 2002, \apj,
  570, 287


\bibitem[Yamamoto et al.(2009)]{Yamamoto2009} 
  Yamamoto, T., et al.\ 2009, The Astron. Telegram, 2297, 1 

\bibitem[Yamamoto et al.(2011a)]{Yamamoto2011a} 
Yamamoto, T., et al.\ 2011, The Astron. Telegram, 3309, 1

\bibitem[Yamamoto et al.(2011b)]{Yamamoto2011b} 
Yamamoto, T., et al.\ 2011, The Astron. Telegram, 3624, 1

\bibitem[Yamamoto et al.(2011c)]{Yamamoto2011pasj} 
Yamamoto, T., et al.\ 2011, \pasj, 63, 751

\bibitem[Yamamoto et al.(2012)]{2012ATel.3856....1Y} 
Yamamoto, T., et al.\ 2012, The Astron. Telegram, 3856, 1

\bibitem[Yamamoto et al.(2014)]{2014PASJ..tmp...44Y} 
Yamamoto, T., et al.\ 2014, \pasj, 66, 59



\end{thebibliography}
\end{document}